\documentclass[aps,twocolumn,showpacs,preprintnumbers,amsmath,amssymb]{revtex4}
\usepackage{graphicx}
\usepackage{dcolumn}
\usepackage{epstopdf}
\usepackage{color}
\usepackage{amssymb}
\usepackage{bm}
\definecolor{dgreen}{cmyk}{1.,0.,1.,0.2}        
\definecolor{orange}{cmyk}{0.,0.353,1.,0.}    

\def\bea {\begin{eqnarray}}
\def\eea {\end{eqnarray}}

\def\be {\begin{equation}}
\def\ee {\end{equation}}

\begin{document}

\title{Method for the Analysis of Forward-Backward Multiplicity Correlations in Heavy-Ion Collisions}

\author{Sudipan De{$^1$}, T. Tarnowsky{$^2$}, T. K. Nayak{$^1$},
  R. P. Scharenberg{$^3$} and B. K. Srivastava{$^3$}} 
\medskip
\affiliation{$^1$Variable Energy Cyclotron Center, Kolkata, India \\
  $^2$National Superconducting Cyclotron Laboratory, Michigan State University, East Lansing, MI-48824, USA \\
  $^3$Department of Physics, Purdue University, West Lafayette, IN-47907, USA }
\bigskip

\date{ \today}
             
\begin{abstract}

In heavy-ion ({\it A-A}) collisions, the correlations among the
particles produced across wide range in rapidity, probe the early
stages of the reaction. The analyses of forward-backward multiplicity 
correlations in these collisions are complicated by several effects, which are absent or
minimized in hadron-hadron collisions. This includes effects, such as
the centrality selection in the {\it A-A} collisions, 
which interfere with the measurement of the dynamical correlations. 
A method, which takes into account the fluctuations in
centrality selection, has been utilized to determine the
forward-backward correlation strength {$b_{\rm corr}$} in {\it
A-A} collisions. This method has been validated by
using the HIJING event generator in case of Au-Au collisions at $\sqrt{s_{NN}}$= 200 GeV and Pb-Pb collisions at $\sqrt{s_{NN}}$= 2.76 TeV.
It is shown that the effect of impact parameter fluctuations is to be
considered properly in order to obtain meaningful results.

\end{abstract}

\pacs{25.75.-q,25.75.Gz,25.75.Nq,12.38.Mh}
\maketitle

\section{Introduction}

The major goals of colliding heavy-ions at relativistic energies are to create a new form of matter, 
called quark-gluon plasma (QGP), and to study its properties. The QGP matter is formed very early in the reaction and it is a major
challenge to experimentally probe this initial stage as majority of the detected particles are emitted at
freeze-out. Correlations, that are produced 
across a wide range in rapidity are thought to reflect the earliest
stages of the heavy-ions collisions, 
free from final state effects~\cite{larry}. 
The study of correlations among particles produced in different rapidity
regions may provide an understanding of the elementary (partonic) interactions which lead to the hadronization. 
Several experiments involving collisions of electrons, muons and protons 
show strong short-range correlations (SRC) over a region of about $\pm$1 unit in rapidity \cite{alner,SRC,aexopoulos}. 
In high-energy nucleon-nucleon collisions ($\sqrt{s} \gg$ 100 GeV) the 
non-single differactive inelastic cross section increases significantly with energy, 
as does the magnitude of the long-range forward-backward multiplicity correlations. 
The component involving the long range correlations in these collisions has been shown to  
increase with the energy \cite{aexopoulos}. These effects can be understood in terms of multiparton interactions \cite{walker2004}. 
In case of heavy-ion collisions, it has been predicted that multiple 
parton interactions would produce long-range forward-backward
multiplicity 
correlations that extend beyond one unit in rapidity, compared to hadron-hadron scattering at the same energy. 
The model based on multipomeron exchanges (Dual Parton Model) predicts the existence of long range correlations 
\cite{capela1,capela2}. In the Color Glass Condensate (CGC) framework,
the correlations of the particles created at 
early stages of the collision can spread over large rapidity
intervals, unlike the particles 
produced at the later stage \cite{larry}. Thus the measurement of the
long range rapidity 
correlations of the produced particle could give information about the space-time dynamics of the early stages of the collisions.  

The measurement of forward-backward (FB) multiplicity correlations in
nucleus-nucleus collisions has been 
studied by the STAR experiment at
RHIC~\cite{star_lrc,tjt2,tjt3,bks1,bks2}. 
These results have generated a great deal of theoretical 
interest~\cite{LRCCGC,HSD,pajares1,Bzdak,Yan1,Gelis,Yan2,Lappi,Bialas,Wit,Broniowski}.

 Forward-backward correlations have been characterized by the
 correlation strength, {$b_{\rm corr}$}, the slope
 extracted from a linear relationship between the average multiplicity
 measured in the backward rapidity hemisphere ($\langle N_{b}
 \rangle$) and the multiplicity in the forward rapidity hemisphere, $N_{f}$. 
This relationship can be expressed as \cite{alner}:
\begin{equation} 
\langle N_{b}(N_{f}) \rangle = a + b_{\rm corr}N_{f}. 
\end{equation}
In this definition, $b_{\rm corr}$ can be
positive or negative with a range of $|b_{\rm corr}| < 1$. This
maximum (minimum) represents total correlation (anti-correlation) of
the produced particles separated in rapidity. $b_{\rm corr} = 0$ is
the limiting case of entirely uncorrelated particle
production. Experimentally, the slope of {$b_{\rm corr}$} in
hadron-hadron experiments is found to be positive \cite{alner}. 
In Eq.(1), the intercept, $a$, is related to the number of uncorrelated particles. 

The correlation strength can also be expressed in terms of the ratio of the
covariance of the forward-backward multiplicity and the variance of
the forward multiplicity. This is done by performing a linear
regression of Eq.(1) and minimizing $\chi^{2}$. Thus, Eq.(1) can be
expressed in terms of the following
calculable average quantities,
\begin{equation}
b_{\rm corr} = \frac{\langle N_{f}N_{b}\rangle - \langle N_{f} \rangle
  \langle N_{b} \rangle }{<N_{f}^{2}>-<N_{f}>^{2}}
= \frac{D_{\rm bf}^{2}} {D_{\rm ff}^{2}},
\end{equation}
where $D_{\rm ff}^{2}$ and $D_{\rm bf}^{2}$ are the forward-forward and backward-forward dispersions. 

The correlations obtained from above expressions can be a combination
of both short and long-ranges.
The short-range correlations (SRC) normally extend over a small range
of pseudorapidity ($|\eta|<1.0$) and are due to various short-range
order effects \cite{alner}. These correlations can arise from various
effects, such as particles produced from cluster decay, 
resonance decay, or jet correlations. 
The particles produced in a single inelastic collision are known to only exhibit SRC~\cite{SRC}.
Long-range correlations (LRC) are correlations that extend over a wide
range in pseudorapidity, beyond $|\eta| > 1.0$. 
The presence of LRC is a violation of short-range order. Short-range
order is expected to hold as long 
as unitarity constraints are neglected~\cite{alner}. 
In the approximation of short-range order, only single scattering can
be considered. Therefore, quantum mechanical probability is not conserved, since it is possible to have multiple scattering terms. 

Recently, FB correlations have been studied extensively with different
model simulations, particularly the 
Color Glass Condensate (CGC)~\cite{Lappi} model and the Color String Percolation
model (CSPM) \cite{pajares2}. 
The CGC provides a QCD based description and predicts the growth of
LRC with collision centrality. 
It is argued that long-range rapidity correlations are due to the
fluctuations of the number of gluons and
can only be created in early times shortly after the collision~\cite{LRCCGC,Lappi}.
In CGC the long range component has the form:
\begin{equation}
b_{\rm corr}=\frac {1}{1+c\alpha_{s}^{2}},
\end{equation}
where $ \alpha_{s}^{2}$ is coupling constant and is related to the
saturation momentum $Q_{s}^{2}$ and $c$ is a constant. From the above
expression, it is observed that as the centrality increases the FB
correlation also increases because $\alpha_{s}^{2}$
decreases~\cite{pajares2}. 
A similar behavior is also obtained in the CSPM
approach. In the CSPM, $b_{\rm corr}$ is expressed in terms of the
string density $\xi$, which is related to the number of strings formed in the collision:
\begin{equation}
b_{\rm corr}=\frac {1}{1+ \frac {d}{(1-e^{-\xi})^{3/2}}},
\end{equation}
which vanishes at low string density and at high density grows to become
$1/(1+d)$, where $d$ is a constant, independent of the density and energy \cite{pajares2}.
The experimental data for Au+Au collision at  $\sqrt{s_{NN}}$= 200 GeV
\cite{star_lrc} show similar trends as predicted by CGC and CSPM. 

FB correlation strength has also been studied in the framework of
wounded nucleon model \cite{Bzdak,wounded}. 
The results are compared to the STAR data \cite{star_lrc} in Au+Au
collisions at $\sqrt{s_{NN}}$= 200 GeV.
It has been concluded that FB correlation strength for central collisions
are due to the fluctuations of 
wounded nucleons at a given centrality bin. Thus it is essential to
control the centrality of the collisions while reporting the
experimental results on correlations.

In the data analysis adopted for the STAR experiment~\cite{star_lrc},
the centrality was defined using the charged particle multiplicity in
the mid rapidity region.  To avoid a self-correlation of the 
results with the window used for centrality definition, a profile
method was used.  In this paper we investigate the profile method to extract the LRC strength in 
heavy-ion collisions and demonstrate its applicability in Pb-Pb
collisions at 
$\sqrt{s_{NN}}$ = 2.76 TeV using HIJING event generator \cite{hijing}.

\section{Analysis Method}

\begin{figure}[thbp]
\centering
 \includegraphics[width=0.5\textwidth]{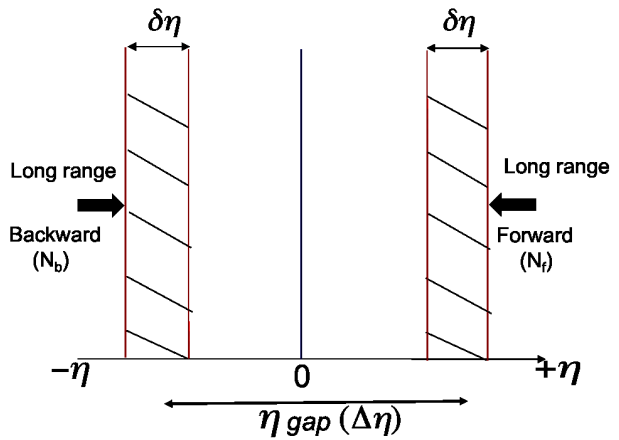} 
\caption{ (Color online) Schematic diagram of the measurement of a forward-backward correlation.}
\label{fig1}
\end{figure}

In a center-of-mass coordinate system, the forward and backward hemispheres 
have been conventionally defined to be opposite to each other, as shown in the 
schematic diagram of figure~\ref{fig1}. $N_f$ and $N_b$ are 
the charged particle multiplicities within the forward and backward
measurement intervals within a width of $\delta\eta$. In our analysis,  a value of $\delta\eta=0.2$ has been 
chosen.  The FB correlations are measured symmetrically around $\eta=0$ with varying rapidity gaps,
designated as $\Delta\eta$, measured from the center of each bin. Thus depending on the available $\eta$ window,
the values of $\Delta\eta = 0.2, 0.4, 0.6, 0.8, 1.0 ....$ are possible. 

In this analysis, data from the 
HIJING event generator have been used, in which the particles are produced 
based on perturbative QCD processes \cite{hijing}. Nearly one million minimum bias 
Au-Au events at $\sqrt{s_{NN}}$=200 GeV and
Pb-Pb events at $\sqrt{s_{NN}}$=2.76 TeV have been generated and used
for the analysis. 
The centrality of the collision is normally designated interms of the
impact parameter of the collision.
In the experiments, it is not possible to determine the impact
parameter directly, hence one uses charged particle multiplicity within a range
of $\eta$, which is not overlapping with the $\eta$ range where the
analysis is performed. This is called reference multiplicity ($N_{\rm ref}$). 
The use of non-overlapping pseudo-rapidity regions, one for the centrality
determination and other ones for FB analysis, avoid bias on the
correlation measurements. 
In the experiments, it is ideal to obtain reference multiplicity from
very forward measurement of charged particles. But if this is not
available, then the centrality can be defined from the central windows
as well.
For example, in the present study, for determining FB correlations in 
$\Delta\eta$ = 0.2, 0.4 and 0.6, reference multiplicity has been obtained within  $0.5<|\eta|<1.0$,
while for $\Delta\eta$ = 0.8 and 1.0 the sum of the multiplicities
from $|\eta|<0.3 $ and $0.8<|\eta|<1.0$ used for centrality determination.
For $\Delta\eta$ = 1.2, 1.4, 1.6 and 1.8,.... the centrality is taken from $|\eta|<0.5$. 
In the correlation analysis, the centrality windows are normally selected
over a range of cross section, which correspond to a range in
reference multiplicity. 
Within a given centrality window, 
the FB multiplicity correlations can be affected by the fluctuations in impact parameter and number of participants.
In order to extract true correlation, it is desirable to control the centrality and minimize the effect of centrality fluctuations.

To calculate the correlation strength as a function of
$\eta$-gap and as a function of centrality, two different method have been discussed. 
In the first method, the quantities such as, 
 $\langle N_{f} \rangle$, $\langle N_{b} \rangle$, $\langle N_{f}^{2} \rangle$ and $\langle N_{f} N_{b} \rangle$, have been obtained by
averaging over the events within a centrality bin, and thereby calculating the dispersions, $D_{\rm ff}^{2}$ and $D_{\rm bf}^{2}$.
This method of event averaging does not take the fluctuation within a
centrality window into account. 
This method is termed as $FB_{average}$ method. 

\begin{figure}[thbp]
\centering
\includegraphics[width=0.50\textwidth,height=3.5in]{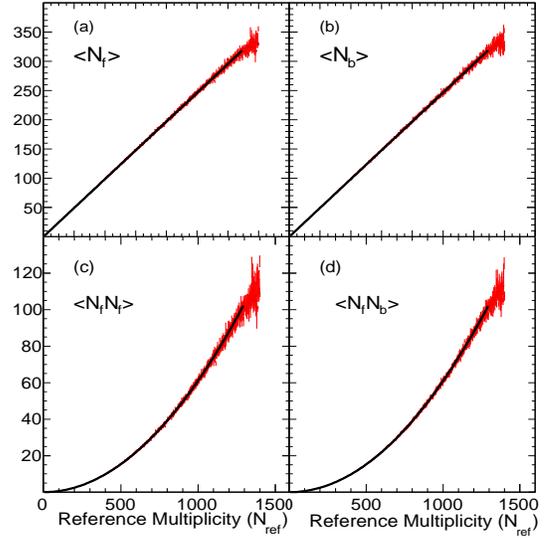} 
\vspace*{-1.5cm} 
\caption{(Color online) Average multiplicities and their products for  Pb-Pb collisions at $\sqrt{s_{NN}}$= 2.76TeV as a function of reference
multiplicity ($N_{\rm ref}$). The upper panels show:
(a) Mean forward charged particle
multiplicity ($\langle N_{f} \rangle$), (b) Mean backward charged multiplicity ($\langle N_{b} \rangle$), 
both fitted with linear polynomial functions. The lower panels show:
(c) $\langle N_{f}N_{f} \rangle $ and (d) $\langle N_{f} N_{b}
\rangle$, both fitted with a second order polynomials.}
\label{fig2}
\end{figure}

\begin{figure}[hbtp]
\centering
\vspace*{-0.2cm}
\includegraphics[width=0.5\textwidth]{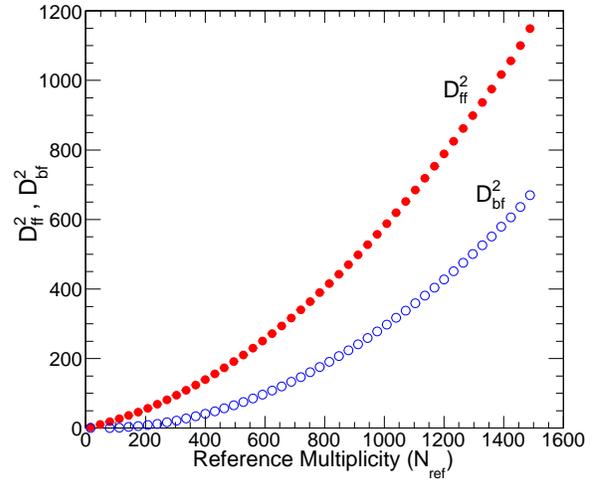} 
\caption{ (Color online) Variance ($D_{\rm ff}^{2}$) and covariance ($D_{\rm bf}^{2}$) as a function of reference multiplicity  
for Pb-Pb collisions at $\sqrt{s_{NN}}$ = 2.76 TeV.
}   
\label{fig3}
\end{figure}

\begin{figure}
\centering
\includegraphics[width=0.55\textwidth]{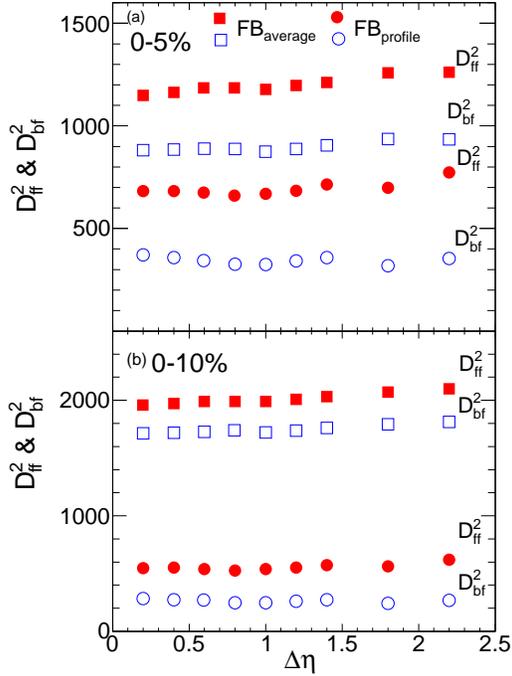}
\vspace*{-1.5cm} 
\caption{(Color online) Comparison of $D_{\rm ff}^{2}$ and $D_{\rm bf}^{2}$ using
FB$_{average}$ and FB$_{profile}$ methods. The results are shown for
0-5\% and 0-10\% centrality for Pb-Pb collisions at $\sqrt{s_{NN}}$
= 2.76 TeV. The centrality is selected using charged particle
multiplicity within a particular pseudo-rapidity window.
}
\label{fig4}
\end{figure}

 In order to eliminate or reduce the effect of the centrality window
on the correlation analysis, a second method, called the profile
method ($FB_{\rm profile}$) has been introduced. 
In this method, the distributions of $\langle N_{f} \rangle$, $\langle N_{b}
\rangle$, 
$\langle N_{f}^{2} \rangle$ and $\langle N_{f} N_{b}
\rangle$,  have been plotted as a function of the reference multiplicities. 
Linear fits to $\langle N_{f} \rangle$, $\langle N_{b} \rangle$ and
second order polynomial fits to 
$\langle N_{f}^{2} \rangle$ and $\langle N_{f} N_{b} \rangle$ have
been made. These distributions, along with the fits are shown in Fig.~\ref{fig2}.
These fit parameters are 
used to extract the $D_{\rm ff}^{2}$ and $D_{\rm bf}^{2}$, binned by
centrality, and normalized by 
the total number of events in each bin. This is shown in
Fig.~\ref{fig3}. 
This method removes the dependence of the FB correlation strength on the width of the centrality bin.
In the next section, results from both the average and profile methods will be presented and compared.

\begin{figure}
\centering
\includegraphics[width=0.55\textwidth]{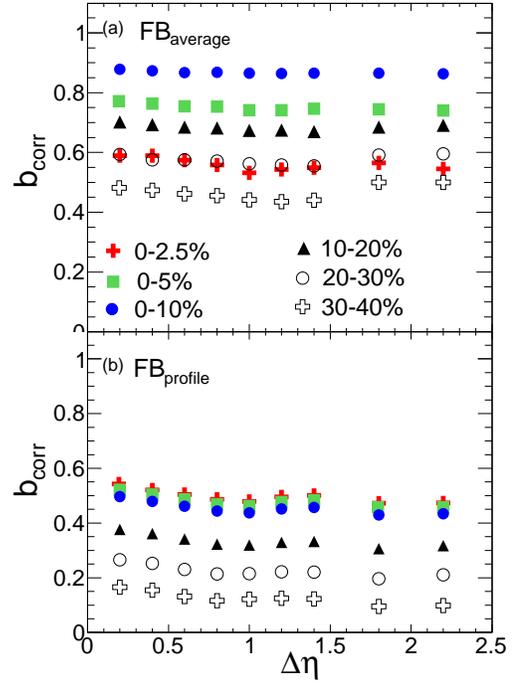} 
\vspace*{-1.5cm} 
\caption{ (Color online) Correlation strength $b_{\rm corr}$ as a function of $\eta$ gap for 5 centrality bins using 
(a) from  $FB_{\rm profile}$ and (b) from $FB_{\rm average}$ for Pb-Pb collisions at $\sqrt{s_{NN}}$ = 2.76 TeV.}
\label{fig5}
\end{figure}

\begin{figure}
\centering
\includegraphics[width=0.55\textwidth]{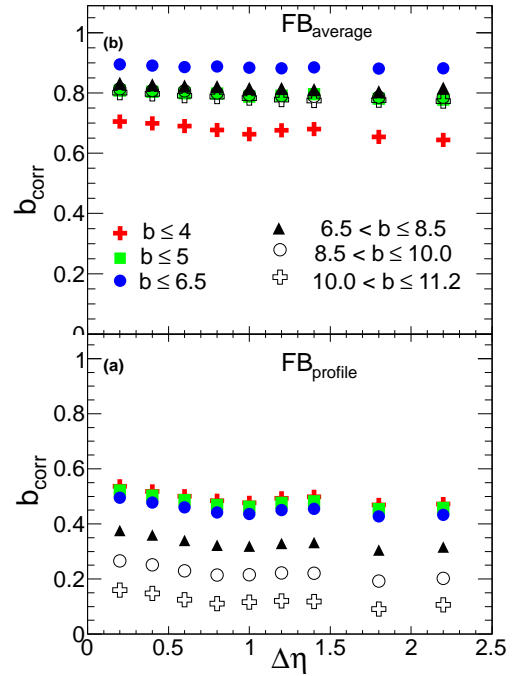} 
\vspace*{-1.5cm} 
\caption{ (Color online) Correlation strength $b_{\rm corr}$ as a function of $\eta$ gap for various impact parameters using 
(a) $FB_{\rm profile}$ and (b) $FB_{\rm average}$ methods for Pb-Pb collisions at $\sqrt{s_{NN}}$ = 2.76 TeV.}
\label{fig6}
\end{figure} 

\begin{figure}
\centering
\includegraphics[width=0.55\textwidth]{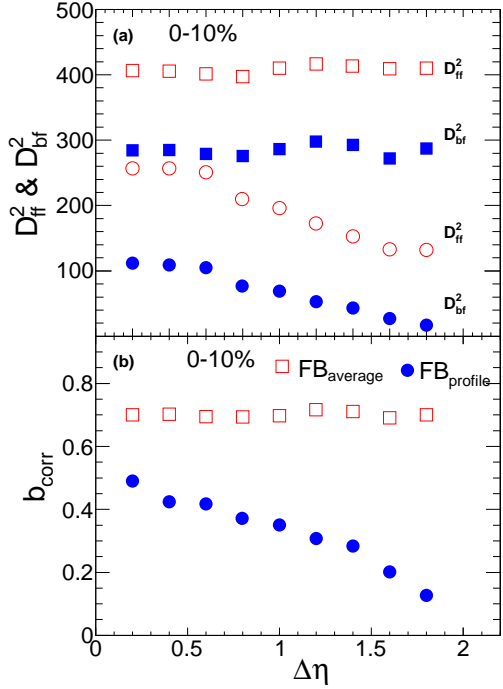} 
\vspace*{-1.5cm} 
\caption{ (Color online) (a).$D_{\rm ff}^{2}$ and $D_{\rm bf}^{2}$ and (b). Correlation strength $b_{\rm corr}$ as a function of $\eta$ gap for 0-10\% centrality from $FB_{\rm profile}$ and (b) method $FB_{\rm average}$ for Au-Au collisions at  $\sqrt{s_{NN}}$ = 200 GeV.}
\label{fig7}
\end{figure} 

\begin{figure}
\centering
\includegraphics[width=0.5\textwidth]{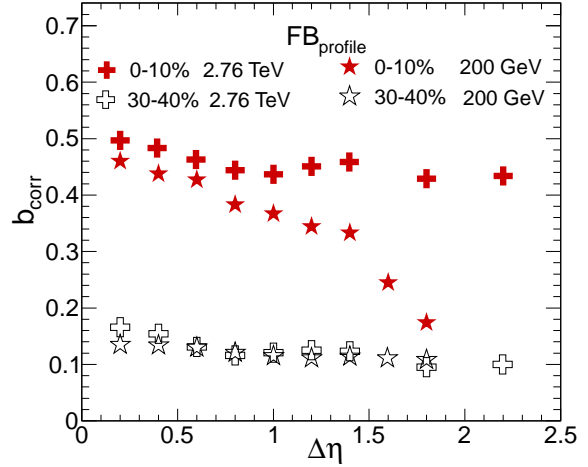} 
\caption{(Color online) Correlation length as a function of rapidity gap  for
 0-10\% and 30-40\% centralities  in case of Au-Au collisions at
$\sqrt{s_{NN}}$ = 200 GeV and  Pb-Pb collisions at $\sqrt{s_{NN}}$ =
2.76 TeV. For central collisions, a clear difference in the
correlation length has been observed.
} 
\label{fig8}
\end{figure} 

\section{Results and Discussion}

The FB correlations have been studied for Pb-Pb collisions at
$\sqrt{s_{NN}}$= 2.76 TeV using the HIJING event generator.
The forward-forward and backward-forward dispersions are calculated
as a function of centrality, within a pseudo-rapidity gap extending up to 2.2 units,
using both average and profile methods.
Figure~\ref{fig4} shows $D_{\rm ff}^{2}$ and $D_{\rm bf}^{2}$ as a
function of $\Delta\eta$ for two overlapping centralities, 0-5\% and
0-10\% of total cross sections. The dispersions remain approximately
constant over the 
rapidity ranges covered. It is observed that $FB_{\rm average}$ yields higher values of both 
 $D_{\rm ff}^{2}$ and $D_{\rm bf}^{2}$ compared to $FB_{\rm profile}$. This is true for both the centrality windows.

It is expected that the correlation strength increases with the
increase of the centrality of the collision. The correlation
strengths, $b_{\rm corr}$, are calculated from the ratios of the
dispersions for six different centrality windows, 0-2.5\%, 0-5\%,
0-10\%, 10-20\%, 20-30\%, and 30-40\% of the cross section. These centralities are
determined from the reference multiplicities as discussed above.
Results from both the methods are presented in Fig.~\ref{fig5}, where
the upper panel shows the values of $b_{\rm corr}$ using $FB_{\rm average}$ method and the lower
panel gives the results for $FB_{\rm profile}$ method. We observe that
$b_{\rm corr}$ for $FB_{\rm average}$ method 
does not follow any regular pattern in terms of centrality selection. 
For example, the $b_{\rm corr}$ is seen to be higher for 
the 0-10\% centrality bin compared to 0-2.5\% and 0-5\% centrality
bin, which is counter intuitive to our expectation.  
This shows that the impact parameter fluctuations are not completely
removed when  $FB_{\rm average}$ method is used.
On the other hand, it can be seen that using
the $FB_{\rm profile}$ method, the values of $b_{\rm
  corr}$, has an increasing trend with the increase of centrality of the
collisions. The correlation strength is highest for 0-2.5\% centrality, as expected.

In order to confirm the above observation, a study using the impact parameter window for centrality selection, rather than
the reference multiplicity, has been made. Results for $b_{\rm corr}$ for various impact parameter selections are shown in 
Figs.~\ref{fig6}(a) and (b),
for the average and profile methods, respectively. The average method
arrives at improper results. In this example, the
larger centrality window yields highest correlation strength, which should not be the case. 
On the other hand, the $FB_{\rm profile}$ method gives similar
results whether centrality selection is made using impact 
parameter or the reference multiplicity. 

A similar study has been made for Au-Au collisions at $\sqrt{s_{NN}}$=
200~GeV using HIJING event generator for top 10\% in central collisions.
The upper panel of Fig.~\ref{fig7} show $D_{\rm ff}^{2}$, $D_{\rm
  bf}^{2}$ and the lower panel shows $b_{\rm corr}$, respectively, for
both the average and profile methods. 
The $FB_{\rm average}$ results yield higher values for  $b_{\rm corr}$
compared to $FB_{\rm profile}$.
The profile results are similar to what had been reported by the STAR Collaboration at RHIC~\cite{star_lrc,tjt3}.

Finally, a comparison of the correlation strengths have been made for 
Au-Au collisions at $\sqrt{s_{NN}}$ = 200 GeV and Pb-Pb collisions at $\sqrt{s_{NN}}$ =
 2.76 TeV using the results from HIJING event generator, and following
 the $FB_{\rm profile}$ method. The results of the study for two
 centrality windows (0-10\% and 30-40\%) are shown in Fig.~\ref
 {fig8}. It is observed that, for the non-central collisions of
 30-40\% cross section, the correlation strengths are very similar.
For central collisions, a
 decreasing trend is observed for Au-Au collision at $\sqrt{s_{NN}}$ =
 200 GeV, whereas for Pb-Pb collisions at $\sqrt{s_{NN}}$ = 2.76 TeV,
 a flatter distribution is observed. This implies a much stronger
 correlation over a broad range in pseudorapidity at the LHC energy compared to those at RHIC.

\section{Summary}

Study of forward-backward multiplicity correlation strengths in hadron-hadron
and heavy-ion collisions provide crucial information towards understanding particle
production mechanisms and represent useful tool for differentiating
different types of reactions and their energy dependence. It has been
observed that the correlations show strong short range correlations
and also extend to much wider separation in
rapidity. In heavy-ion collisions, the correlation strengths are expected to increase with
increase of the beam energy as well as centrality of the
collision. Within a given centrality window, the fluctuations in the
impact parameter or the number of participants lead to multiplicity
fluctuations which affect the accurate determination of correlation
strength. It is therefore needed to control the centrality of the
collisions while performing the correlation analysis. 

In this manuscript, two different methods, the average method and the
profile method, have been presented to study the
forward-backward multiplicity correlations in heavy-ion collisions as
a function centrality. It is observed that in the $FB_{\rm average}$
method, the correlation strength does not follow any pattern as a
function of centrality window. This reflects the impact parameter
fluctuation due to finite centrality bin width.
The second method, $FB_{\rm profile}$, has been introduced, which
properly takes care of the effects due to finite centrality bin
width. Appropriate centrality dependence has been observed in going
from peripheral to central method. A comparison of the correlation
strengths have been made for Au-Au collision at $\sqrt{s_{NN}}$= 200 GeV and
Pb-Pb collisions at $\sqrt{s_{NN}}$= 2.76 TeV using the data from
HIJING event generator. It has been observed that the correlation
strengths are higher for higher energy collision. The correlation
strengths decrease as a function of the rapidity gap. This decrease is
much slower at LHC energy compared to that of the RHIC energies. 
The $FB_{\rm profile}$ method can be used to study the FB correlation
strength as a function of centrality in the Pb-Pb collisions at LHC. 
 As is shown in this work, along with the correlation strength ($b_{\rm corr}$), it is essential to show the behavior of both the forward-forward ($D_{ff}^2$) and backward-forward ($D_{bf}^2$) dispersions as a function of pseudo-rapidity gap ($\Delta\eta$) for different centrality classes. This will allow to make a direct comparison of experimental data with theoretical models, such as, CGC and CSPM. 
       
\section{Acknowledgements}     
 This research was supported by the Office of Nuclear Physics within
 the U.S. Department of Energy  Office of Science under Grant No. DE-FG02-88ER40412.
 S.D and T.N were supported by the Department of Atomic Energy, Government of India.



\begin{thebibliography}{00}

\bibitem{larry} Y. V. Kovchegov, E. Levin, and L. McLerran,
Phys. Rev. C{\bf 63}, 024903 (2001).

\bibitem{alner}
G. J. Alner {\it et al.}, 
Phys. Rep. {\bf 154}, 247 (1987).

\bibitem{SRC} A. Capella, and J. Tran Thanh van, 
Z. Phys. C{\bf 18}, 85 (1983).

\bibitem{aexopoulos}
T. Alexopoulos {\it et al.}, 
Phys. Lett. B{\bf 353}, 155 (1995).

\bibitem{walker2004}
W.D. Walker, 
Phys. Rev. D{\bf 69}, 034007 (2004).

\bibitem{capela1} A. Capella and A. Krzywicki, 
Phys. Rev. D{\bf 18}, 120 (1978).

\bibitem{capela2} A. Capella {\it et al.},
Phys. Rep. {\bf 236}, 225 (1994).

\bibitem{star_lrc} B. I. Abelev {\it et al.}  (STAR Collaboration),
Phys. Rev. Lett. {\bf 103}, 172301 (2009).

\bibitem{tjt2} T. Tarnowsky (STAR Collaboration),
Int. J. Mod. Phys. E {\bf 16}, 3363 (2007).

\bibitem{tjt3} T. Tarnowsky (STAR Collaboration),
Proceedings of Science (CPOD07) 019,2007, nucl-ex/0711.1175.

\bibitem{bks1} B. K. Srivastava (STAR Collaboration), 
Int. J. Mod. Phys. E{\bf 16}, 3371 (2007).

\bibitem{bks2} B. K. Srivastava (STAR Collaboration), 
Eur. Phys. J A{\bf 31}, 3862 (2007).

\bibitem{LRCCGC}
N. Armesto, L. McLerran, and C. Pajares,  
Nucl. Phys. A{\bf 781}, 201 (2007).

\bibitem{HSD} V. P. Konchakovski, M. Hauer, G. Torrieri, M. I. Gorenstein, and E. L. Bratkovskaya,
Phys. Rev. C{\bf 79}, 034910 (2009).

\bibitem{pajares1} P. Brogueira, J. Dias de Deus, and C. Pajares,
Phys. Lett. B{\bf 675}, 308 (2009).

\bibitem{Bzdak} A. Bzdak, 
Phys. Rev. C{\bf 80}, 024906 (2009).

\bibitem{Yan1} 
Yu-Liang Yan et al., 
Phys. Rev. C{\bf 79}, 054902 (2009). 

\bibitem{Gelis}  
F. Gellis, T. Lappi, and L. McLerran, 
Nucl. Phys. A{\bf 828}, 149 (2009).

\bibitem{Yan2} 
Yu-Liang Yan et al., 
Phys. Rev. C{\bf 81}, 044914 (2010). 

\bibitem{Lappi} T. Lappi, and L. McLerran, 
Nucl. Phys. A{\bf 832}, 330 (2010). 

\bibitem{Bialas} A. Bialas, and K. Zalewski, 
Phys. Lett. B{\bf 698}, 416 (2011).

\bibitem{Wit} K. Fialkowski, and R. Wit, nucl-th/1203.3671.

\bibitem{Broniowski} A. Olszewski, and W. Broniowski, nucl-th/1303.5280.

\bibitem{pajares2} C. Pajares,
Nucl. Phys. B{\bf 854}, 125 (2011).

\bibitem{wounded} A. Bialas, M. Bleszynski, and W. Czyz, 
Nucl. Phys. B{\bf 111}, 461 (1976). 

\bibitem{hijing} X. -N. Wang, and M. Gyulassy,  
Phys. Rev. D{\bf 44}, 3501 (1991).
\end{thebibliography}
\end{document}